\documentclass[12pt,a4paper,final]{article}
\usepackage[utf8]{inputenc}
\usepackage{amsmath}
\usepackage[english]{babel}
\usepackage{amsfonts}
\usepackage{amssymb}
\usepackage{setspace}
\usepackage{makeidx}
\usepackage{graphicx}
\usepackage[left=2.5cm, right=2.5cm]{geometry}
\addto{\captionsenglish}{}
\date{}
\author{$FAICAL\quad BARZI$\\\small faical.barzi@edu.uiz.ac.ma\\\\\small $^1$Centre Regional des Métiers de l'Education et de la Formation\\\small CRMEF-Marrakech-Safi, Morocco\\\small $^2$Ibn Zohr University-Agadir, Morocco}
\title{\Huge A look into the possibility of negative mass}

\begin{document}
\maketitle

\begin{abstract}
We investigate the possibility of negative mass particles and whether physics as we know it today allows or excludes their existence and what properties they might have in the light these laws. We show that nothing prohibit their existence, in the contrary, for a more symmetric world their existence is essential. In relativistic quantum mechanics, the negative mass bispinor in the rest frame takes a different form than the positive mass bispinor.

\end{abstract}

\section{Introduction}

\paragraph{}Negative mass is a an extension of the habitual concept of mass\cite{concept1}\cite{concept2}. Bondi\cite{Bondi} first considered this extension in the context of general relativity where it has a particular status different from Newtonian mechanics where the action-reaction principal constrains its meaning. Bonnor\cite{Bonnor} reported an earlier suggestion in the nineteenth century and concluded that a universe made up of negative mass is comprehensible. It is clear that no contradictions arise from introducing this extension\cite{thermo}.  Usually we tend to associate mass with the quantity of matter a given system has, as in chemistry, or in the relativistic limit to the energy content of the system and thus positivity of mass seems natural to admit. However, there are many levels to look at the concept of mass. Firstly, we know that there are different types of mass, particularly the inertial and gravitational masses. The inertial mass measures the inertia of a system i.e. its resistance to a change of its state of motion and it appears in Newton's second law of dynamics, it is a proportionality coefficient between the force (excitation) applied to a system and its acceleration(response), in this respect the inertial mass is \textit{an impedance}, a more general concept in physics defined as :

\begin{equation}\label{key}
Impedance=\frac{Excitation}{Response}
\end{equation}
\paragraph{}A negative inertial mass then means that acceleration and the force have opposite directions and does not depend rigorously on the quantity of matter in the system. Assuming the equivalence principal to hold then the ratio of inertial and gravitational masses is universal, therefore gravitational mass could also be negative. The mass-energy equivalence expressed by Einstein equation $E=mc^2$, does not restrict the sign of mass to be positive since this equation generalizes to $E=\sqrt{\textbf{p}^2_{/\mathcal{R}}c^2+m^2c^4}$ in a given inertial frame of reference $\mathcal{R}$. In the rest frame ($\textbf{p}^2=0$) it reads $E=|m|c^2$. In this same respect, particle-antiparticle production and annihilation thresholds are not changed even if we suppose that antiparticles have negative mass. Some authors\cite{Hamm}\cite{attempt1}\cite{attempt2}\cite{attempt3} tried ad'hoc modifications of the expression of the non-relativistic kinetic energy $E_c=\frac{1}{2}mv^2_{\mathcal{/R}}$ by multiplying by a sign factor, forgetting the relativistic origin of this expression. Nevertheless, these authors produced some interesting classes of particles with properties "mimicking" in some respect the behavior of dark matter and dark energy. Although the equivalence principal holds to a great accuracy for matter its validity for anti-matter hasn't yet acquired sufficient proof. A promising tests for the equivalence principal at CERN, namely, the AEGIS\cite{AEGIS1}\cite{AEGIS2}\cite{AEGIS3} experiment propose the measurement of the Earth gravitational acceleration  $\overline{g}$ of anti-hydrogen by a free fall through a moiré deflectometer, which apart from testing the equivalence principal will reveal gravitational properties of anti-matter. In quantum field theory, We know that requirement for gauge invariance prohibits a mass term in the Lagrangian volume density, Like the case of the photon field. The Lagrangian density (eq.\ref{lagrange}) for a massive field $\Psi$ contains instead a Yukawa interaction term with the Higgs field $\Phi$, which after the  electroweak symmetry breaking through the Higgs mechanism generates a mass term for $\Psi$. A priori then mass is just a Yukawa  coupling constant, and nothing prevent it from being negative.

\begin{align}
\mathcal{L}=...+g\overline{\Psi}\Psi\Phi+...
\label{lagrange}
\end{align}
the symmetry between matter an antimatter dictated by the $CPT$ invariance if rigorously true, concern only inertial masses and leave the gravitational masses unrestricted. Furthermore whether negative mass particles are antiparticles is not likely, given the status of the CPT invariance, but they could exist as a new exotic class of mass conjugate particles to the ordinary matter. There is numerous macroscopic and microscopic physical situations where an effective negative mass shows up, these effective masses are due to the interaction of the system with its environment which causes the \textit{bare mass} to be transformed to the effective mass or in the language of QFT a \textit{normalized mass}. Moreover, in their foundational paper\cite{Bohr} about the measurability of electromagnetic fields (EMF), Bohr and Rosenfeld, centered their argument on a key observation, that a measuring apparatus necessary has some external charges that would generate additional EMF that would render the measurement impossible unless another distribution of charge is crafted to cancel or screen the EMF created by the measuring apparatus itself and this is only possible because the Electromagnetic interaction can be either repulsive or attractive, in other words there are positive and negative electric charges. In analogy with the electromagnetism, the measurability of gravitational fields would be therefore impossible unless there are negative masses because a positive mass allows only a repulsive character and no screening is possible. The existence of gravitational negative mass would then strengthen the symmetry between the two fields. Lastly we can draw a parallel between the existence of negative mass particles and magnetic monopoles, for a complete symmetry between electrical and magnetic phenomena we would like to these monopoles to exist. Likewise for more symmetric spectrum of particles, negative mass is needed.
\section{Preliminaries: Effective negative mass}
\subsection{A gas bubble in a dense fluid} 
Given a gas bubble of density $\rho$ and volume $V$ submerged in a static fluid of density $\rho_f>\rho$. The bubble is submitted to a net upward force:
\begin{align}
P=-(\rho_f-\rho)Vg
\end{align}
The bubble has then an effective negative mass $m_{eff}=-(\rho_f-\rho)V<0$, whereas the real or bare mass of the bubble is $m=\rho V>0$. The effective mass result from the interaction between the bubble and its surrounding fluid, it captures the action of the static fluid so that we could ignore its presence and consider only the bubble in "free fall". 
\subsection{An electron in a crystal}
Given an electron in a crystal, it occupies a level in a given band. The state of the electron is then specified by the crystal vector momentum $\hbar \textbf{k}$, the spin $s$ and the band index $n$. Its mean velocity, denoted $\textbf{v}_n(\textbf{k})$ is given by:
\begin{align}
\textbf{v}_n(\textbf{k})=\frac{1}{\hbar}\nabla_{\textbf{k}} E_n(\textbf{k})
\end{align}

Within the semi-classical model\cite{crystal1}, let the band energy $E_n(\textbf{k})$ have its maximum at $\textbf{k}_0$, then for sufficiently symmetric crystal near the maximum the energy can be written in the neighborhood of $\textbf{k}_0$:
\begin{align}
 E_n(\textbf{k})=E_n(\textbf{k}_0)-C(\textbf{k}-\textbf{k}_0)^2+O((\textbf{k}-\textbf{k}_0)^3) 
\label{velo}
\end{align}
with $C$ a positive constant, $C>0$. we introduce the constant mass $m_*$ such that, $C=\frac{\hbar^2}{2 m_*}$. Now using equation (\ref{velo}), the mean velocity of the electron occupying a level $\textbf{k}$ near $\textbf{k}_0$ is:
\begin{align}
\textbf{v}_n(\textbf{k})=-\frac{\hbar(\textbf{k}-\textbf{k}_0)}{m_*} 
\end{align}
Thus the mean acceleration of of the electron is:
\begin{align}
\textbf{a}_n(\textbf{k})=\frac{d\textbf{v}_n(\textbf{k})}{dt}=-\frac{\hbar}{m_*}\frac{d\textbf{k}}{dt}
\end{align}

Now the motion of the electron in a spatially slow varying electromagnetic field $(\textbf{E},\textbf{B})$, is determined in the semi-classical model by the equation:
\begin{align}
&
\hbar\frac{d\textbf{k}}{dt}=-e(\textbf{E}+\frac{\textbf{v}}{c}\times\textbf{B})\\
&
-m_*\textbf{a}_n(\textbf{k})=-e(\textbf{E}+\frac{\textbf{v}}{c}\times\textbf{B})
\end{align}
Therefore near the energy maximum the electron behaves as if it has effective negative mass $-m_*<0$.
\section{Negative mass in classical mechanics}
\subsection{Non-gravitational interaction}
\paragraph{}Let a particle with a negative mass $m$ subjected to a non-gravitational force $\overrightarrow{F}$, From the second law of Newtonian dynamics applied in a frame of reference $\mathcal{R}$ we have:
\begin{equation}
\overrightarrow{F}=\frac{\overrightarrow{dp}}{dt} |_{\mathcal{R}}
\end{equation}
\paragraph{}Thus the particle of negative mass is always accelerated in the opposite direction to the force, surprisingly any attempt to stop a negative mass particle by pushing it away just increase its acceleration, for instance negative mass particle don't bounce of walls or potential barriers but tend to penetrate them and the stronger the barrier the faster the penetration.

\paragraph{}Consider a barrier whose threshold force is $F_{yield}$ i.e the force at which it will yield to penetration. A negative mass particle $m$ normally incident on the barrier with an initial velocity $v_0$ will subject the barrier at time $t>0$ to a force $\overrightarrow{F(t)}$:
\begin{align}
\overrightarrow{F(t)}=\frac{\gamma m\overrightarrow{v}_{|\mathcal{R}}}{\tau}=\frac{\overrightarrow{p(t)}_{|\mathcal{R}}}{\tau}
\end{align}

Where $\tau$ is a time scale of momentum and energy transfer from the particle to the barrier and $\gamma=\frac{1}{\sqrt{1-\frac{v^2}{c^2}}}$, Then the momentum transfer toward the barrier verify:

\begin{align}
&\frac{\overrightarrow{dp}}{dt}_{|\mathcal{R}}=\frac{\overrightarrow{p}}{\tau}\\
&\frac{d(\gamma m v)}{dt}_{|\mathcal{R}}=\frac{\gamma m v}{\tau}\\
&v(t)_{|\mathcal{R}}=\sqrt{\frac{\gamma_0^2v_0^2e^{\frac{2t}{\tau}}}{1+\gamma_0^2\frac{v_0^2}{c^2}e^{\frac{2t}{\tau}}}}\\
&F(t)=\frac{|m|}{\tau}\gamma_0v_0e^{\frac{t}{\tau}}
\end{align}
We see that the force applied on the barrier has no limit: $t\rightarrow\infty$, $F(t\rightarrow\infty)\rightarrow\infty$.

\begin{figure}[h]
	\centering
	\begin{minipage}{.5\textwidth}
		\centering
		\includegraphics[width=.75\linewidth]{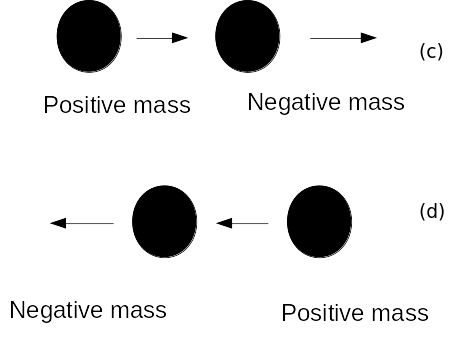}
	\end{minipage}%
	\begin{minipage}{.5\textwidth}
		\centering
		\includegraphics[width=.85\linewidth]{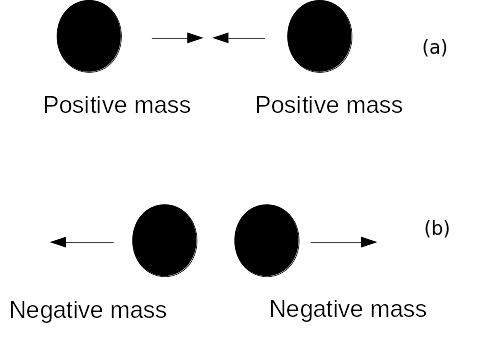}

	\end{minipage}
\caption{\small (c) and (d) Negative and positive mass chase each other: Unilateral electrical attraction or repulsion.}
\label{fig:test1}
\end{figure}

\paragraph{}Therefore the barrier will yield at the time $T_{yield}$ such that:
\begin{eqnarray}
&F_{yield}=\frac{|m|}{\tau}\gamma_0v_0e^{\frac{T_{yield}}{\tau}}\\
&T_{yield}=\tau\ln(\frac{\tau F_{yield}}{|m|\gamma_0v_0})
\end{eqnarray}
The more rigid the barrier, the larger $F_{yield}$, but also the smaller $\tau$ will be i.e the faster the energy transfer happens. So we have $\tau F_{yield}\approx const$. We have thus:
\begin{eqnarray}
T_{yield}\propto \tau \propto \frac{1}{F_{yield}}
\end{eqnarray}
\paragraph{}Which proves that the stronger the barrier the faster the penetration, for instance if the barrier is infinitely rigid it becomes transparent to any negative mass projectile. We see also that the time to yield decreases only logarithmically with the magnitude of the negative mass, so a small negative mass has as much penetration power as a much larger one. It is then clear that classical negative mass projectiles are unstoppable which make the the ultimate anti-armor weapon. 
\paragraph{}Consider a charged negative mass and let $e$ be its charge. Under the influence of an electromagnetic field ($\overrightarrow{E}$,$\overrightarrow{B}$) the particle motion is given by:
\begin{eqnarray}
&\frac{d\overrightarrow{v}}{dt}_{|\mathcal{R}}=\frac{e}{m}(\overrightarrow{E}+\overrightarrow{v}_{|\mathcal{R}}\times \overrightarrow{B})
\end{eqnarray}
\paragraph{}Since this equation depends on the ratio $\frac{e}{m}$, Then a negatively charged negative mass particle behaves in a electromagnetic field identically to a positively charged positive mass particle, and also a positively charged negative mass behaves identically to a negatively charged positive mass. It thus seems plausible to consider only one sign for the electric charge and both signs for the mass instead of the hitherto accepted picture of one sign for mass and both signs for charge. The commonly stated rule \textit{"like charges repel and opposite charges attract"} assumes that the mass is always positive. If the mass were negative we could not say anything meaningful concerning charges alone, since the very definition of what the sign of a given charge is depends on its behavior in an electric field which in turn depends on the sign of the mass. furthermore gravitational force is no longer attractive only, a competition between the two interactions occurs for large masses. We could have instead the rule assuming that electromagnetic interaction dominates\textit{"opposite charges attract(repel) and like charges repel(attract) if they have positive(negative) masses. Otherwise, negative(positive) mass \textbf{chases} (see figure\ref{fig:test1}) positive(negative) mass if they are like(opposite) charges"}.

\paragraph{}These novel situations ((c) and (d) in figure \ref{fig:test1}) are not yet observed in nature but they complete the symmetry of possible interactions between two particles, it seems strange why we should admit that only mutual repulsion or mutual attraction exist. Unilateral repulsion and unilateral attraction should also exist in a nature that like so much to be as symmetric as possible. A pair of negative and positive mass particles will accelerate toward the speed of light\cite{Bondi}.


\subsection{Gravitational interaction}
\paragraph{}Let $m$ be a negative mass particle its gravitational field $\overrightarrow{g}$ at a  distance r is:
\begin{equation}
\overrightarrow{g}=G\frac{|m|}{r^2}\overrightarrow{u}
\label{u}
\end{equation}
Where the unit vector $\overrightarrow{u}$ points away from the negative mass particle.
\paragraph{}Thus, if we the equivalence principal holds a negative mass repels gravitationally any other mass, positive or negative, a same conclusion is reach from general relativistic considerations\cite{schwarchild}. Black holes of positive mass would correspond to white holes of negative mass. This would explain why there are no macroscopic objects made exclusively of electrically neutral negative-mass matter since they can not stick together. The only possibility is a single elementary negative mass particles. If a negative mass particle touches the earth's surface it will penetrate it as we saw above exiting from the other side with great velocity. There seems to be a confusion in many authored papers\cite{misc1}\cite{schwarchild} about the coalescence of negative mass, and using this argument against its possibility on the ground that no macroscopic negative mass object has never been observed.

\paragraph{}Consider a system of two non-relativistic particles with masses $m_1$ and $m_2$ which can take both signs and charges $q_1$ and $q_2$ respectively, initially separated by a distance $r=r_0$. The total potential energy reads:

\begin{eqnarray}
U=\frac{-4\pi\epsilon G  m_1m_2+q_1q_2}{4\pi \epsilon r}
\label{pot}
\end{eqnarray}

\paragraph{}Consider the particular case, $m_1=m>0$ and $m_2=m_-<0$ and charges $q_1=q_+$ and $q_2=q_-$ respectively, initially separated by a distance $r=r_0$. Equation (\ref{pot}) reads:
\begin{eqnarray}
U=\frac{4\pi\epsilon G  m|m_-|+q_+q_-}{4\pi \epsilon r}
\end{eqnarray}
According to the rule above $m_-$ will chase $m$ if $q_+=q_-$ and the other way around if  $q_+=-q_-$. Let's rewrite $q_+=\lambda q_-=q$, where $\lambda=\pm 1$. Then:
\begin{eqnarray}
U=\frac{4\pi\epsilon Gm|m_-|+\lambda q^2}{4\pi \epsilon r}
\end{eqnarray}
Let $\mu =\frac{mm_-}{m+m_-}$ be the reduced mass of the system, then the equation of relative motion is:
\begin{eqnarray}
&\mu\frac{d^2r}{dt^2}=-\nabla U(r)\\
&\implies\mu\frac{d^2r}{dt^2}=\frac{ 4\pi\epsilon Gm|m_-|+\lambda q^2}{4\pi \epsilon r^2}\\
&\implies\frac{d^2r}{dt^2}=-\frac{G(m+m_-)}{r^2}+\frac{\lambda q^2}{4\pi \epsilon\mu r^2}\\
&\implies\frac{d^2r}{dt^2}=-\frac{G(m-|m_-|)}{ r^2}+\frac{\lambda q^2}{4\pi\epsilon \mu r^2}\label{case}
\end{eqnarray} 
We see that if $m=|m_-|$ then $\mu\longrightarrow\infty$ and  $\frac{d^2r}{dt^2}=0\implies r=v_ot+r_o$ where $v_o$ and $r_o$ are the initial relative velocity and the initial separation respectively. In the interesting case $m>>|m_-|$ and $q_+=0$, then $\mu\approx m_-<0$. equation (\ref{case}) becomes:

\begin{align}
\frac{d^2r}{dt^2}\approx -\frac{Gm}{r^2}
\end{align}

Which correspond to an attraction, thus the Earth, say, will attracts a small negative mass particle the same way it would do for positive mass particle due the large mass difference. The gravitational force applied by the Earth on a the negative mass particle is:
\begin{eqnarray}
\textbf{F}=-\frac{Gm|m_-|}{r^2}\textbf{u}
\end{eqnarray}

Where $\textbf{u}$ is directed towards the earth (see eq.(\ref{u})). Therefore the acceleration of the particle is towards the earth owing to its negative mass. Note that the gravitational potential energy $U=\frac{G  m|m_-|}{r}$, is positive even though the particle is attracted to the Earth. The rule that \textit{a negative potential energy means attraction and a positive potential energy imply repulsion} is no more valid since it does depend on the sign of the mass, for negative mass the opposite apply i.e. \textit{a negative potential energy means repulsion and a positive potential energy imply attraction}

\paragraph{}Now many negative mass particles of the same charge electrically attract each other although they gravitationally repel each other but they could nevertheless stick together because electrical forces are much stronger than gravitational ones, then this macroscopic object will be charged. Its total potential energy $U$ for $N$ identical negative mass particles of charge $e$ can be written:
\begin{equation}
U=\sum_{i<j}^{N}\frac{e^2-4\pi\epsilon Gm^2}{4\pi \epsilon r_{ij}}
\end{equation}
Or, for a charge and mass densities $\rho(r)$ and $\mu(r)$:
\begin{eqnarray}
U=\frac{1}{2}\int d^3rd^3r' \frac{\rho(r)\rho(r')-4\pi\epsilon G\mu(r)\mu(r')}{4\pi \epsilon |r-r'|}
\end{eqnarray}
Let's consider the case of a sphere of radius $R$ with uniform densities, Then if $Q$ and $M<0$ are the total charge and mass respectively:
\begin{eqnarray}
U=\frac{3}{20\pi \epsilon R}(Q^2-4\pi\epsilon GM^2)>0
\end{eqnarray} 

The kinetic energy of the system is:
\begin{eqnarray}
&T=\int \:(\sqrt{c^4dm^2+(dp({\small \overrightarrow{r}}))^2c^2}-c^2|dm|)>0\\\\
&T=\int \:(\gamma -1)c^2|dm|>0
\end{eqnarray}
Which remains positive, in the non-relativistic limit $T$ becomes:
\begin{eqnarray}
T=-\frac{3M}{8\pi R^3}\int d^3r\:v^2(r)_{|\mathcal{R}}>0
\end{eqnarray}
According to the Virial theorem which must be modified  for the case at hand the system's time averaged kinetic energy is:
\begin{eqnarray}
<T>_t=\textbf{+}\frac{1}{2}<U>_t=\frac{3}{40\pi \epsilon R}(Q^2-4\pi\epsilon GM^2)>0
\end{eqnarray}
So the kinetic energy of the system is positive as it should be. In cosmology the discovery that the universe is accelerating that prompted a huge interest in negative mass\cite{cosmo1}\cite{cosmo2}. Previously, Einstein introduced his cosmological constant to get static solutions to his equation whose role was to compensate the gravitational attraction of positive mass matter, it provided a repulsive force against gravitation mimicking negative mass behavior \cite{Hamm}.

\section{Negative mass in Thermodynamics}
\begin{figure}[h]
	\centering
	\includegraphics[width=0.7\linewidth]{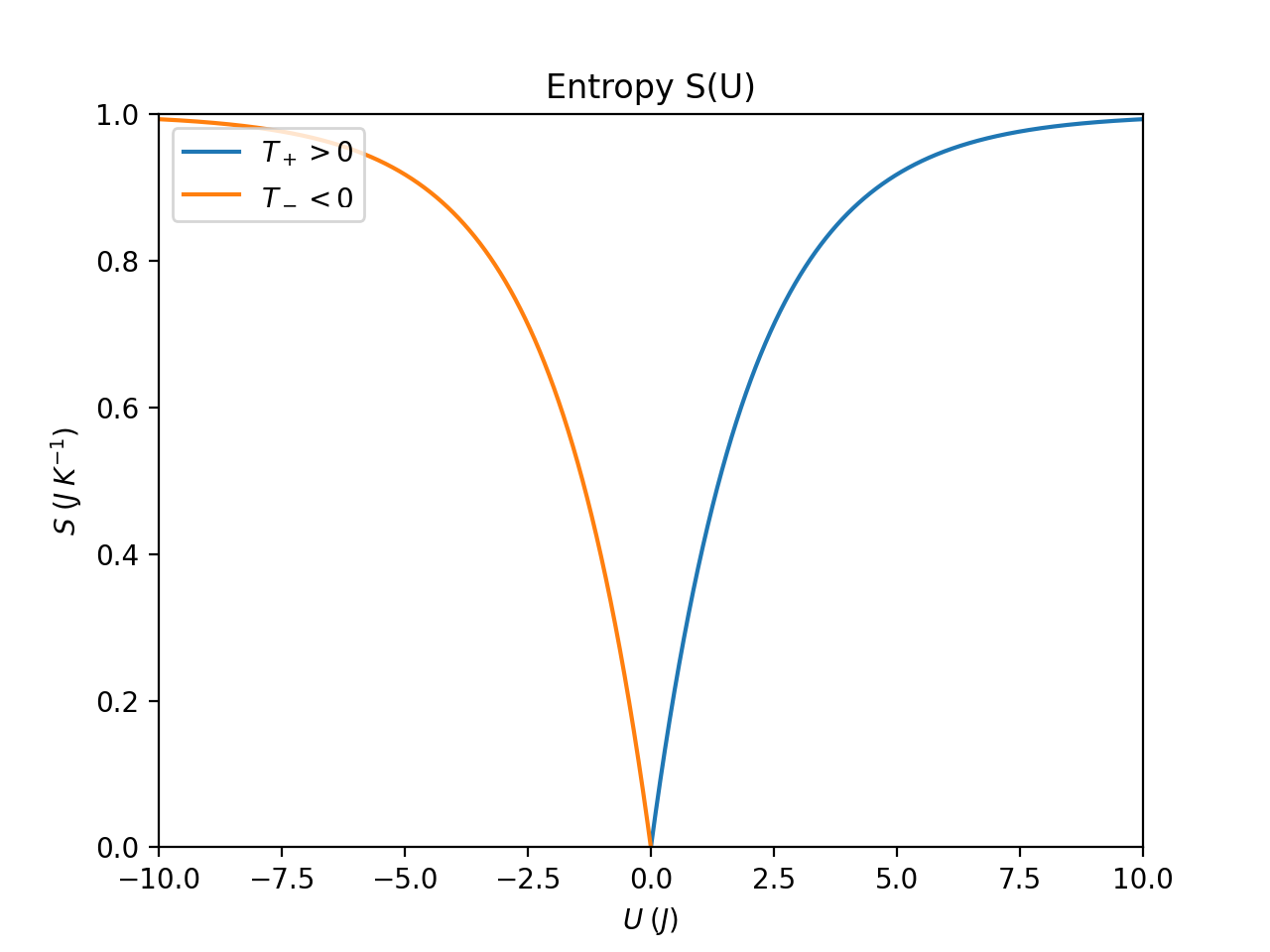}
	\caption{Entropy as function of internal energy for positive and negative temperature thermodynamic system.}
	\label{fig:entropy}
\end{figure}

In a thermodynamic system containing positive and negative inertial mass particles the question arises to what thermodynamic parameters should describe its equilibrium state. In \cite{thermo} the interesting conclusion is reached that a classical ideal gas at thermodynamic equilibrium with only negative mass constituents could only have negative temperature as $\frac{1}{2}m_-v_{/\mathbf{R}}^2=\frac{3}{2}k_BT$, and coexistence with positive mass which require a positive temperature is impossible precisely because no equilibrium state would be reached for the combined system. Terletsky\cite{thermo2} claimed a violation of the second law of thermodynamics of a process where a positive mass particle emits a negative mass particle, gaining thus internal energy and jumping to an exited state. But it is well-known that entropy $S$ is a increasing function of internal energy $U$ as $\frac{dS}{dU}=\frac{1}{T}$ for positive mass system (FIG.\ref{fig:entropy}) , or at least a decrease in entropy of the positive mass will be compensated by an increase in the entropy of the negative mass such that the overall balance is an increase of entropy since the number of particles increased.

The individual partition function of a classical ideal gas of negative mass particles in a unit volume is known to be:
\begin{equation}\label{key}
	Z_-= \frac{(2\pi m_-kT_-)}{\sqrt{h}} ^{\frac{3}{2}}	
\end{equation}

Where $m_-<0$. No doubt this partition function is only defined if with negative mass we have a negative temperature as well $T_-<0$, such that $m_-T_->0$. The individual partition function for a positive mass ideal gas is of the same form $Z_+= \frac{(2\pi m_+kT_+)}{\sqrt{h}} ^{\frac{3}{2}}$. Let's combine $N_-$ negative mass particle and $N_+$ positive mass particle. The total partition function $Z$ of the composed system  reads:
\begin{align}\label{key}
& Z=\frac{Z_-^{N_-}}{N_-!}\frac{Z_+^{N_+}}{N_+!}\\
&Z=	 \frac{(2\pi m_-kT)}{\sqrt{h}N_-!} ^{\frac{3N_-}{2}} \frac{(2\pi m_+kT)}{\sqrt{h}N_+!} ^{\frac{3N_+}{2}}
\end{align}
We see the ambiguity in the sign of non zero temperature $T$, which must be negative and positive at the same time. The only possibility is $T=0\:K$, but at such temperatures quantum effects may dominate for bosons which require further investigation. Thus except at absolute zero temperature, negative mass and positive mass are thermodynamically isolated from each other. This of course could explain the lack of observation of negative mass systems in nature since absolute negative temperature situations are not common\cite{thermo3}.

\section{Negative mass in Quantum mechanics}
We look at negative mass particles in non-relativistic and relativistic quantum mechanics respectively.
\subsection{Negative Mass in Galilean relativity}
The laws of non relativistic quantum mechanics are covariant under Galilean space and time transformations, that is, under a translation of space origin, a translation of time origin, a rotation of space axes and a constant relative velocity between two inertial frame of reference. These transformations constitute the Galilean group. The subgroup of infinitesimal transformations has the form:
\begin{align}
	\mathbf{ r_2}=\mathbf{r_1}-\mathbf{\delta r} \quad t_2=t_1-\delta t
\end{align}
Where,
\begin{align}
\mathbf{\delta r}= \mathbf{\Xi}+ \mathbf{\Omega} \wedge \mathbf{r}+ \mathbf{v}\delta t
\end{align}
and $\mathbf{\Xi}$, $\mathbf{\Omega}$ and $\mathbf{v}$ are infinitesimal space displacement, rotation and velocity respectively. To gain more information about the structure of this group we look at the composition law of two infinitesimal transformations. But first we note that quantum mechanics has an arbitrariness where all states can be changed by a common phase factor, we associate this arbitrariness with a unitary transformation generated by the unity operator in addition of the generators of the Galilean group. Thus an infinitesimal unitary transformation $U=1+iG$ of a quantum state can be written as:
\begin{equation}\label{key}
G=\frac{1}{\hbar}(\mathbf{\Xi}\: \mathbf{P}+\mathbf{ \Omega}\: \mathbf{J}+\mathbf{ v}\: \mathbf{N} -\delta t\: H) + \phi \:I	
\end{equation}

Where $\mathbf{P}$, $\mathbf{J}$, $\mathbf{N}$ and $H$ are the generators of the Galilean group. $\phi$ is an infinitesimal phase angle. Let $U_1$ and $U_2$ be two infinitesimal unitary transformations and consider the composition:
\begin{equation}\label{key}
	U_{[12]}=U_1U_2-U_2U_1
\end{equation}
Let $\Xi_{12}$, $\Omega_{12}$, $v_{12}$ and $\phi_{12}$ be the parameters associated with $U_{[12]}$. We find that a general form for $\phi_{12}$ is:
\begin{equation}\label{key}
	\phi_{12}=M(\mathbf{\Xi_1 v_2-\Xi_2 v_1})
\end{equation}
This comes from the observation that $\phi$ is a scalar and the only scalars we can make from the parameters of $U_1$ and $U_2$ are these above. The remarkable thing is that the constant M turns out to be the total mass of the quantum system\cite{schwinger} such that the boost generators $\mathbf{N}$ are given by:
\begin{equation}\label{key}
	\mathbf{N}=\mathbf{P}\: t-M \: \mathbf{R}
\end{equation}
Where $\mathbf{R}$ is the position operator. The generators of the Galilean group are constants of the motion, therefore $\frac{d\mathbf{N}}{dt}=\mathbf{0}$ and the commutator $[\mathbf{N},H]=0$:
\begin{align}\label{key}
&	\frac{d\mathbf{N}}{dt}=\frac{\partial \mathbf{N}}{\partial t}+[\mathbf{N},H]=0\\
&\implies	\mathbf{P}=M\frac{d\mathbf{R}}{dt}
\end{align}
Which shows that $M$ is the total mass. Now apart from being a constant, no restrictions are put on the sign of $M$. Non-relativistic quantum mechanics alone has nothing to say about the sign of the mass it admits both. This, of course leads through Hamilton's equations of motion to the non-relativistic form of the Hamiltonian, $H=\frac{P^2}{2M}+ H_{int}$. It's the relativistic form that imposes the sign on $M$ as already pointed out in the introduction.
\subsection{Schrodinger equation}
We consider two mass quanta $m_a=\lambda_a |m_a|$ and $m_a=\lambda_b |m_b|$, the Schrodinger equation describing their gravitational interaction is:
\begin{equation}\label{key}
H\Psi(r_a,r_b)=E\Psi(r_a,r_b)
\end{equation} 
The Hamiltonian $H$ of the two quanta system is written:
\begin{equation}\label{key}
H=\frac{P^2_a}{2|m_a|}+\frac{P^2_b}{2|m_b|}+V_{ab}
\end{equation}
where $P_a=-i\hbar\nabla_a$ and $P_b=-i\hbar\nabla_b$ are the momenta of the two mass quanta and $V_{ab}$ is the gravitational potential of interaction, $V_{ab}=V(r_a-r_b)\propto-\frac{\lambda_a\lambda_b}{|r_a-r_b|}$. Note that the expressions for classical non-relativistic kinetic energies are approximations of the relativistic expression which is independent of the sign of the mass ($E_c=(\gamma-1)|m|c^2$).
\paragraph{}Let $r=r_a-r_b$ and $R=\frac{|m_a|r_a+|m_b|r_b}{|m_a|+|m_b|}$, then the problem can be separated into two independent motions, the motion of the center of mass and the relative motion referenced by $R$ and $r$ respectively:
\begin{equation}
[\frac{-\hbar^2}{2M}\nabla_R^2+\frac{-\hbar^2}{2\overline{\mu}}\nabla_r^2+V(r)]\Psi(R,r)=E\Psi(R,r)	
\end{equation}
Where we define the total mass $M$ and the reduced mass $\overline{\mu}$:
\begin{eqnarray}
&M=|m_a|+|m_b| \\
&\overline{\mu}=\frac{|m_a||m_b|}{|m_a|+|m_b|}
\end{eqnarray}
 
 Since the two motions are independent the wave function $\Psi$ and the energy $E$ can be written:
\begin{eqnarray}
	&\Psi(R,r)=\xi(R)\zeta(r)\\
	&E=E_R+E_r
\end{eqnarray}
Thus we obtain two equations:\\
\begin{eqnarray}
&\frac{-\hbar^2}{2M}\nabla_R^2\xi(R)=E_R\xi(R)\\             
&[\frac{-\hbar^2}{2\overline{\mu}}\nabla_r^2+V(r)]\zeta(r)=E_r\zeta(r)
\end{eqnarray}
We see that depending on the sign of the mass quanta $m_a$ and $m_b$, we have three cases:

\begin{enumerate}
	\item $\lambda_a=\pm1$ and $\lambda_b=\pm1$: The interaction is always attractive, this is the normal case.
	\item $\lambda_a=\pm1$ and $\lambda_b=\mp1$: The interaction is always repulsive. 
	
	
\end{enumerate}	
The center of mass wave $\xi(R)$ is just a plane wave:
\begin{equation}
\xi(R)=C\:exp({\frac{i}{\hbar }\sqrt{2ME_R}R})
\end{equation}
The quantum mechanical treatment through the Schrödinger equation is not influenced by the sign of the masses in the same way as does the classical treatment through Newton second law of dynamics. Quantum mechanically the problem is similar to the interaction of two electrical charges by coulomb force.
\subsection{Klein-Gordan equation}
The Klein-Gordan equation (KGE) was the first relativistic quantum equation proposed. It reads in natural units:
\begin{equation}
	(\partial_\mu^2-m^2)\psi(x)=0
\end{equation}
Because of its second order nature, it admits negative energy solutions and there is a difficulty in interpreting the density of probability $\rho(x)$ and the current density $\overrightarrow{j(x)}$ derived from it. They read:
\begin{eqnarray}
	&\rho=\frac{i}{2\lambda m}(\psi^*\partial_0\psi-\psi\partial_0\psi^*)\\
	&\textbf{j}=-\frac{i}{2\lambda m}(\psi^*\nabla\psi-\psi\nabla\psi^*)
\end{eqnarray}  
where $\lambda$ is now the sign of the mass. Let's look at the probability and current densities of a free plane wave solution of (KGE):
\begin{equation}
	\psi(x)=\frac{1}{\sqrt{V}}exp[-i(k_0t-\textbf{k.x})]
\end{equation}
Where $k_0=\pm\sqrt{\textbf{k}^2+m^2}=\pm\frac{\lambda m}{\sqrt{1-\textbf{v}^2}}$. We find:
\begin{eqnarray}
	&\rho_{free}=\frac{k_0}{\lambda m}\\
	&\textbf{j}_{free}=\frac{\textbf{k}}{\lambda m}
\end{eqnarray}
Then $\rho_{free}$ is negative for a negative energy free particle with positive mass and we can not interpret it as probability density as it could be done for a positive energy free particle with positive mass. Therefore, it appears tempting to consider a negative mass for negative energy states in order to retain the positivity of $\rho$, and state that \textit{A positive mass and energy particle with a given momentum behaves the same as a negative mass and energy particle with the opposite momentum.} However the TCP theorem imposes that anti-particles must have the same inertial mass as their particles, therefore these negative energy and mass candidate particles are not the observed anti-particles but must be different entities.
\subsection{Dirac equation}
The Dirac equation being first order is sensitive to the mass sign. For a bi-spinor $u$ of positive energy $+E$ and  momentum $\textbf{p}$ such that the state $\psi=u$ $exp[-i(Et-\textbf{p.x})]$ it reads:
\begin{equation}{\label{dirac1}}
	Eu=(\mathbf{\alpha}.\mathbf{p}+\beta m)u
\end{equation}
Where $\alpha$ is a matrix vector  $\alpha=(\alpha^1\: \alpha^2\: \alpha^3)$. (\ref{dirac1}) gives the following system of coupled equations of the components of the bi-spinor $u(p)$ :

\begin{align}	
&	(E-m)u_1=p_3u_3+(p_1-ip_2)u_)\\
&	(E-m)u_2=(p_1+ip_2)u_3-p_3u_4\\
&	(E+m)u_1=p_3u_1+(p_1-ip_2)u_2\\
&	(E+m)u_4=(p_1+ip_2)u_1-p_3u_2
\end{align}

Only two components, $u_1$ and $u_2$ or $u_3$ and $u_4$ are independent. In terms of $u_3$ and $u_4$, The system has thus two independent solutions:
\begin{align}
	 \psi_{+\frac{1}{2}}=e^{[-i(Et-\textbf{p.x})]}\left( \begin{array}{ccc}
		1  \\
		0 \\
		\frac{p_3}{E+m} \\
		\frac{p_1+ip_2}{E+m}  \end{array} \right) &\quad \psi_{-\frac{1}{2}}=e^{[-i(Et-\textbf{p.x})]} \left( \begin{array}{ccc}
		0  \\
		1 \\
    	\frac{p_1-ip_2}{E+m} \\
		-\frac{p_3}{E+m} \end{array} \right)
\end{align}

Which represent the two spin  free states of the electron with positive energy $E$ and momentum $\textbf{p}$ . For negative energy solutions they can not be physical because they are in contradiction of matter stability : a hydrogen atom would decay to one of these state in less than $10^{-10}s$\cite{dyson}. Dirac supplied us with the prescription that all negative energy states are normally filled with one electron and no transition can thus be made to these states due the Pauli exclusion principal, However if a negative energy state of energy $-E$ and momentum $\textbf{-p}$ becomes empty (by knocking out the electron that occupies it), the hole left behind must appear as a positive energy state with energy $+E$,  momentum $\textbf{p}$ and carrying a positive charge. This is the positron state.

\paragraph{}If we transform the positive mass $m$ to a negative mass $-m$, then $\textbf{p}\longrightarrow-\textbf{p}$. The state of positive mass transforms to u$\longrightarrow \tilde{u}$ we get:

\begin{eqnarray}{\label{dirac2}}
&E\tilde{u}=(-\mathbf{\alpha.p}-\beta m)\tilde{u}
\end{eqnarray}
We want nature to treat negative mass states $v$ on equal footing with positive mass, thus for a negative mass state $v$, the Dirac equation should read:
\begin{eqnarray}{\label{dirac3}}
&Ev=(\mathbf{\alpha.p}-\beta m)v
\end{eqnarray}
Thus we must have:
\begin{equation}{\label{dirac4}}
	 v=M\tilde{u}
\end{equation} 
Where $M$ is \textit{the mass inversion} transformation. Of course, inverting the mass twice we should regain the original sign thus $M^2=\mathbb{I}$ . From (\ref{dirac2}), matrix multiplying both sides from the left by $M$, we get:
\begin{eqnarray}
	&EM\tilde{u}=(-M\mathbf{\alpha.p}-M\beta m)\tilde{u}
	\label{dirac5}
\end{eqnarray}
replacing (\ref{dirac4}) in (\ref{dirac3}):
\begin{eqnarray}
&EM\tilde{u}=(\mathbf{\alpha.p}-\beta m)M\tilde{u}\\
&\implies EM\tilde{u}=(\alpha M\mathbf{.p}-\beta M m)\tilde{u}
\label{dirac6}
\end{eqnarray}
Comparing (\ref{dirac5}) and (\ref{dirac6}), we must have:
\begin{equation}
M\alpha^k=-\alpha^kM,\quad M\beta =\beta M,\quad M^2=\mathbb{I}
\end{equation}
Where $k=1,2$,$3$ and $\alpha=(\alpha^1\: \alpha^2\: \alpha^3)$. Thus $M$ anti-commutes with all $\alpha^k$'s and commutes with $\beta$, in contrast with charge conjugation transformation $C$ which commutes with all the $\alpha^k$ and anti-commutes with $\beta$\cite{dyson}.
\subsection{Poincaré group}
\paragraph{}Ultimately, The study of the kind of particles permitted to exist in nature comes to the careful investigation of the representations of symmetry groups that we impose on it. The Poincaré group is the minimum symmetry group of particle physics and it seems a priori that it allows negative mass states. We investigate the behavior of these states as representations of this group. There are four classes of particle states singled out by the Poincaré group, the class of real mass particles, the class of massless particles, the vacuum class and the class of imaginary mass particles. Only the first and last classes are of interest to us.
\paragraph{} Let's look at first massive particle of real mass $m$, they satisfy the on-shell relation:
\begin{equation}
	p^2_\mu=-m^2
\end{equation}
Where $p_\mu=(\textbf{k},ik_0)$ is the 4-momentum of the particle, $p_0=\pm\sqrt{m^2+\textbf{$p^2$}}$=$\pm \omega_p$.
\paragraph{}Let $\left|\textbf{p},s,\pm\right\rangle$ be a complete set of states vectors for the reducible unitary representation of massive particles. The $s$ symbol represents the spin degree of freedom and $\pm$ the positive or negative energies. Let $\left|\psi\right\rangle$ be the state vector of a massive particle of negative mass, we have:
\begin{equation}
	\psi(\textbf{p})_{s,\pm}=\left\langle\textbf{p},s,\pm |\psi\right\rangle	
\end{equation}
	Under a Lorentz transformation $\Lambda$,  $\psi(\textbf{p})_{s,\pm}$ transforms in the following way:
	\begin{equation}
			\psi'(\textbf{p})_{s,\pm}=D_{ss'}(\textbf{p})	\psi(\Lambda^{-1}\textbf{p})_{s,\pm}
	\end{equation}
	Where $D_{ss'}(\textbf{p})$ is a representation of the little group $SO(3)$\cite{rep}. For an infinitesimal pure rotation with angles $\theta=(\theta_x,\theta_y,\theta_z)$ we can write:
	\begin{equation}
			\psi'(\textbf{p})_{s,\pm}=(1+i \textbf{$\theta$J}) \psi(\textbf{p})_{s,\pm}
	\end{equation} 	
	Where \textbf{$J$} are representations of the generators of rotations:
	\begin{equation}
	\textbf{J}=-i\textbf{p}\times\frac{\partial}{\partial\textbf{p}}+\textbf{S}
	\end{equation}
	Where \textbf{S} is the spin generator. These generators do not depend on mass at all, so negative and positive mass particles will transform the same way under pure rotations. For infinitesimal pure boosts with rapidities $\xi=(\xi_x, \xi_y, \xi_z)$ we can write:
		\begin{equation}
	\psi'(\textbf{p})_{s,\pm}=(1-i \textbf{$\xi$K}) \psi(\textbf{p})_{s,\pm}
	\end{equation} 
	Where \textbf{$K$} are representations of the generators of boosts:	
		\begin{equation}
	\textbf{K}=-p_0(-i\frac{\partial}{\partial\textbf{p}}-\frac{\textbf{p$\times$S}}{\omega_p(m+\omega_p)})
	\end{equation}
	We see that $\textbf{K}$ depends on mass and we expect the behavior of positive and negative mass to be significantly different. Consider a normalized positive energy and negative mass ($m<0$) free solution of the Dirac equation with $s=\frac{1}{2}$ (omitting spacetime dependence):
		\begin{equation}
	\psi^-(\textbf{p})_{\frac{1}{2},+}=\sqrt{\frac{\omega_p+m}{2|m|\omega_p}}
	  \left( \begin{array}{ccc}
	1  \\
	0 \\
	\frac{p_3}{\omega_p+m} \\
	\frac{p_1+ip_2}{\omega_p+m}  \end{array} \right)
	\label{freestate}
	\end{equation}
	Where the superscript minus sign  indicates a negative mass and we adopt the normalization, a particle per volume $\frac{\omega_p}{|m|}$, i.e. \begin{align}
	\psi^-(\textbf{p})_{\frac{1}{2},+}^\dagger\psi^-(\textbf{p})_{\frac{1}{2},+}=\frac{\omega_p}{|m|}
	\label{norm}
	\end{align}.
	 In the rest frame of the negative mass particle, this state is apparently not normalized because of the divergence of $\frac{1}{\sqrt{\omega_p+m}}$  when $p\rightarrow0$.
%
	 Take a free positive energy and negative mass particle moving along the $x$-axis. Equation ($\ref{freestate}$) reads:
		\begin{equation}
	\psi^-(\textbf{p})_{\frac{1}{2},+}=\sqrt{\frac{\omega_p+m}{2|m|\omega_p}}
	\left( \begin{array}{ccc}
	1  \\
	0 \\
	0 \\
	\frac{p_1}{\omega_p+m}  \end{array} \right)
	\end{equation}
	
	Where $\omega_p=\sqrt{m^2+\textbf{p}_1^2}$.
	\paragraph{}As said above, we can not put $p_1=0$, Let's then compute the limit $p_1\longrightarrow0$, suppose that $p_1=\epsilon$, where $\epsilon<<|m|$:
\begin{align}
	\omega_p\longrightarrow \sqrt{m^2+\epsilon^2}\longrightarrow |m|(1+\frac{\epsilon^2}{2m^2})	\\
	\omega_p+m\longrightarrow |m|\frac{\epsilon^2}{2m^2}
\end{align}

Then, in the rest frame the negative mass state reads :
	\begin{equation*}
\psi^-(\textbf{p}_1\longrightarrow 0)_{\frac{1}{2},+}=\frac{\epsilon}{\sqrt{4|m|^3(1+\frac{\epsilon^2}{2m^2})}}
\left( \begin{array}{ccc}
\epsilon  \\
0 \\
0 \\
\frac{2|m|}{\epsilon}   \end{array} \right)\overset{\epsilon\rightarrow0}{\longrightarrow}\frac{1}{\sqrt{4|m|^3}}
\left( \begin{array}{ccc}
0 \\
0 \\
0 \\
2|m|  \end{array} \right)
\end{equation*}
\begin{equation}
\longrightarrow\frac{1}{\sqrt{|m|}}
\left( \begin{array}{ccc}
0 \\
0 \\
0 \\
1  \end{array} \right)
\end{equation}

Thus the limit exist and negative mass particles have  \textit{a rest frame}, since the state in the rest frame is normalizable according to equation (\ref{norm}). In contrast to a positive mass state, ($m>0$) in its rest frame we have:

	\begin{equation}
\psi^+(\textbf{p}_1=0)_{\frac{1}{2},+}=\frac{1}{\sqrt{|m|}}
\left( \begin{array}{ccc}
1  \\
0 \\
0 \\
0  \end{array} \right)
\end{equation}
Thus the two states are orthogonal in same rest frame \begin{equation}\label{key}
\psi^-(\textbf{p}_1=0)_{\frac{1}{2},+}^\dagger\psi^+(\textbf{p}_1=0)_{\frac{1}{2},+}=0
\end{equation}
This is a Lorentz-invariant statement valid in any inertial frame of reference:
\begin{equation}\label{key}
\psi^-(\textbf{p})_{\frac{1}{2},+}^\dagger\psi^+(\textbf{p})_{\frac{1}{2},+}=0
\end{equation}

\subsection{Negative mass and the range of an interaction}
Since the work of H.Yukawa\cite{hediki} on the nucleon-nucleon interactions, the range of an interaction is measured by the mass of the messenger particle exchanged during the interaction. The heavier the exchanged particle is the shorter the range. The Yukawa potential expresses this relation in the form of $V$ as:
\begin{equation}\label{key}
	V \propto \frac{e^{-m r}}{r}
\end{equation}

\begin{figure}[h]
	\centering
	\includegraphics[width=0.7\linewidth]{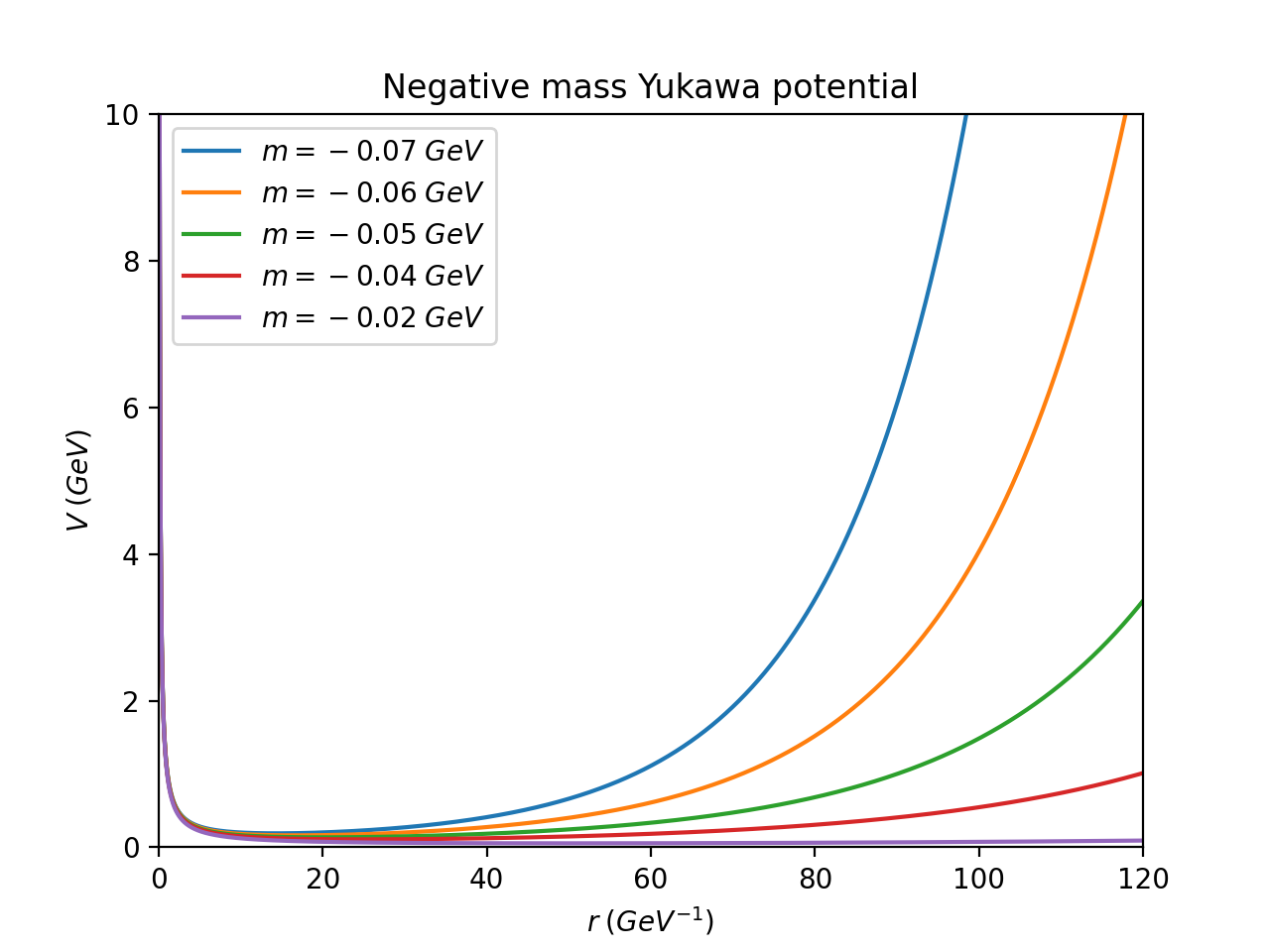}
	\label{yukawa}
	\caption{Yukawa potential with a negative mass exchanged particle, the exponential increase dominates only beyond a certain distance.}
	\label{fig:yukawa2}
\end{figure}

Where the $m$ is  the mass of the exchanged particle. The electromagnetic interaction has no well-defined range except from a geometrical weakening because the photon is massless. A negative mass force carrying particle ($m<0$) would imply also an interaction with no definite range and an exponential strengthening. At present we don't know of such interaction except that the strong nuclear force in a crude approximation does increase with  distance but somewhat linearly. If however $m$ is small the exponential increase is dominated by the geometrical weakening as seen from (FIG.\ref{fig:yukawa2}), and only beyond a certain separation $r_{min}$ does the exponential strengthening dominate.

The "turning point" separation $r_{min}$ is given by $r_{min}=-\frac{1}{m}$. Now in order for an interaction with a negative mass carrier to appear weakening with distance up to a 1 m, $|m|$ must be less than $0.2\;\mu eV$.

\newpage

\section{Conclusion}
It is clear from the above that the laws of physics don't exclude in any way the existence of negative mass particles. We showed that their incapacity to cling together may explain why there are no electrically neutral macroscopic negative mass bodies. Relativistic quantum mechanics does not exclude their existence too. Anti-matter could have a negative mass whose effects would only show up in an gravitational interaction with positive mass matter.

\end{document}